# Accelerator and Reactor Neutrino Oscillation Experiments in a Simple Three-Generation Framework

G.L. Fogli[1], E. Lisi[1,2] and G. Scioscia[1]

[1]*Dipartimento di Fisica and Istituto Nazionale di Fisica Nucleare, Bari, Italy*
[2]*Institute for Advanced Study, Princeton, NJ 08540*

## Abstract

We present a new approach to the analysis of neutrino oscillation experiments, in the one mass-scale limit of the three-generation scheme. In this framework we reanalyze and recombine the most constraining accelerator and reactor data, in order to draw precise bounds in the new parameter space. We consider our graphical representations as particularly suited to show the interplay among the different oscillation channels. Within the same framework, the discovery potential of future short and long baseline experiments is also investigated, in the light of both the recent signal from the LSND experiment and the atmospheric neutrino anomaly.

# 1 Introduction

Neutrino flavor oscillations [1] are an elegant tool to search, at low energy, for physics beyond the Standard Model of electroweak interactions, and hopefully to gain information on very high energy scales via the see-saw approach to $\nu$ masses [2].

Laboratory neutrino beams from accelerator and reactors are, in principle, perfectly suited to pursue oscillation searches, since the beam spectrum and/or the path length can be controlled to a large extent. However, apart from a tentative indication recently reported by the LSND experiment [3], at present the only possible evidences for neutrino oscillations originate from natural beams: solar neutrinos and atmospheric neutrinos (see the reviews [4] and [5], respectively).

This situation could rapidly change in the near future: data from running laboratory experiments and the developement of new, powerful detectors placed at large distances from the neutrino beam sources (long baseline experiments) will make it possible to push the experimental sensitivity to neutrino oscillations toward the domain of the present "atmospheric beams," and possibly beyond.

Such a rich phenomenology cannot be handled consistently in a two-generation approximation and requires a transition to three-neutrino schemes. This paper intends to present a comprehensive analysis of the present data from accelerator and reactor experiments in the simplest limit of the general three-flavor neutrino mixing scenario, known as "one mass-scale dominance" [6, 7]. In the same scenario, the discovery potential of future short and long baseline experiments will be investigated.

Several phenomenological analyses of neutrino oscillation data have already been published in the one mass-scale dominance limit, with particular attention to accelerator and reactor data [6, 7, 8], to atmospheric data [9, 10, 11, 12], or both [13, 14, 15], the subdominant mass scale usually being assumed to solve, in some way, the solar neutrino problem. In such a scheme, attempts to fit the "17 KeV neutrino" [16] or $\sim 200$ MeV neutrinos [17] have also been made.

Very recently, this approach has been applied in Refs. [18, 19] to get sketchy indications from the tentative LSND signal, and in Refs. [20, 21] to develop a more systematic and detailed analysis of recent atmospheric, accelerator and reactor data, as well as of future experiments.

As far as laboratory beam experiments are concerned, our goal is more ambitious. In fact, we have completely reanalyzed *ab initio* the most constraining published data, in order to combine all the experimental results, and obtain statistically significant bounds in the new parameter space. As far as we know, this is the first such attempt since 1985 [22], and is one distinguishing aspect of our work. Moreover, new and useful graphical representations are introduced, which not only show easily and neatly the interplay between the different oscillation channels probed by present and future experiments, but are also able to display results in both the small and in the large mixing regime. Atmospheric neutrino data also fit in this scheme, but their rich phenomenology (see [13, 14, 23] and references therein) deserves a separate work [24], and here they will be considered only qualitatively. Solar neutrino data are not considered in this work.

This paper is presented in the following way. Present and future accelerator/reactor experiments are reviewed in Sec. 2. In Sec. 3, the basics of the one mass dominance formalism are recalled, and the graphical representations that will be used in the analysis are introduced. The complete reanalysis of the present data is presented in Sec. 4, together with its implications on atmospheric neutrinos. In Sec. 5, the discovery potential of future short baseline experiments is discussed, in the light of the recent LSND results. Similarly, future long baseline experiments are discussed in the light of the atmospheric neutrino anomaly. We summarize our main results in Sec. 6. The Appendix is meant to document our reanalysis of accelerator and reactor data thoroughly.

## 2 Survey of Present and Future Experiments

Neutrino oscillation experiments with laboratory beams can be divided into several classes, according to the source (accelerator or reactor), the final/initial flavor comparison (flavor appearance or disappearance), or the specific oscillation channel probed. Another useful (although more subjective) division is into long and short baseline experiments; in the following, we will call "short" a source-to-detector baseline $L \lesssim 1$ km.

We anticipate that the possible CP violation effects are unobservable in our framework, so $\nu$ and $\bar{\nu}$ will not be necessarily distinguished and the channels will be simply labeled by the detected flavors, as $\mu \leftrightarrow \tau$, sometimes using a single-arrow notation (as $\mu \to \tau$) to distinguish the initial flavor.

### 2.1 Short Baseline Experiments

All laboratory experiments performed so far fall into this category (for a review see Ref. [25]). The most constraining results in the various flavor oscillation channels have been obtained by the following experiments: E776 (accelerator, $\mu \to e$) [26], E531 (accelerator, $\mu \to \tau$ and $e \to \tau$) [27], CDHSW (accelerator, $\mu \to \mu$) [28], Krasnoyarsk (reactor, $e \to e$) [29] and Bugey (reactor, $e \to e$) [30]. The combination of the latter two experiments improves the results previously obtained at the Gösgen reactor ($e \to e$) [31]. None of these experiments has found any evidence for neutrino oscillations.

Currently, there are four running short baseline experiments: LSND [3] and KARMEN [32], exploring the $\mu \to e$ oscillation channel, and CHORUS and NOMAD [33] at CERN, exploring the $\mu \to \tau$ channel. The first two have presented preliminary results [3, 32], which are somewhat in contradiction. In fact, the LSND Collaboration has reported a tentative oscillation signal, not confirmed—but not entirely ruled out—by KARMEN. Moreover, a more conservative treatment of background and cuts for the LSND data could make the signal vanish [34]. We can only wait for more robust (dis)proofs of the LSND evidence for oscillations. As far as CHORUS and NOMAD are concerned, they are expected to present results in about one year. As for the E531 experiment, they are also expected to use the small (less than 1%) $e$-flavor contamination in the beam to explore, as a bonus, the $e \to \tau$ channel, although with much less sensitivity than for $\mu \to \tau$ searches.

To conclude the short baseline survey, we mention three future projects which are being vigorously pursued: the reactor (disappearance) experiments at Chooz [35] and San Onofre [36] (see also [37]), and the accelerator experiment E803 (COSMOS) at FNAL [38]. COSMOS will be similar to NOMAD, but with an expected increase in sensitivity by at least one order of magnitude.

## 2.2 Long Baseline Experiments

Although the idea of studying neutrino oscillations with long baseline experiments is not new [39] (for a more recent discussion see [40]), such projects are still at the proposal stage, since they require a dedicated and costly orientation of the beam, and large-volume detectors. The interest in such experiments has, however, increased in the last few years, since they would probe the region of oscillation parameters suggested by the atmospheric flavor anomaly (on this subject, see [13, 14] and references therein). At present, various possibilities are being debated in Europe, U.S.A. and Japan.

The current European proposals are to send a beam from CERN to Gran Sasso ($L = 732$ km) in Italy—ICARUS being the most probable detector [41, 42]—and/or to the NESTOR detector in Greece ($L = 1676$ km) [43]. Both would explore $\mu \to e$ oscillations; ICARUS is also planned to detect $\tau$'s in the final state.

The American project P-875 (MINOS) [44] intends to send a $\nu_\mu$ beam from FNAL to the Soudan mine ($L = 730$ km), and explore, with different techniques, three channels: $\mu \to e, \mu, \tau$. That will provide powerful cross-checks of possible oscillation signals. A somewhat parallel experiment is also being considered: P-822 [38], using the same facilities and an upgraded version of the existing Soudan 2 detector. Another long-baseline project is BNL E889 at Brookhaven [38], a purely $\mu \to \mu$ disappearance experiment with one close (1 km) and two far (3, 24 km) detectors.

The only Japanese proposal, if funded, is likely to be the first to operate in the next few years: sending a neutrino beam from KEK-PS to SuperKamiokande ($L = 250$ km) [45]. This will allow the experiment to operate in the $\mu \to e$ channel and, perhaps, in the disappearance mode $\mu \to \mu$. The low energy of the beam ($\sim 2$ GeV) will prevent detection of $\mu \to \tau$ flavor oscillations.

We are not concerned here with the so-called extremely long accelerator neutrino (ELAN) oscillation searches [40, 46], which could be sensitive to matter enhancement effects on a several thousand km path length through the Earth's crust. At present, none of them seems to be really considered as feasible. Thus, the vacuum oscillation formalism will be used throughout this paper.

## 2.3 Performances of the Experiments

In the two-flavor approach, the oscillation probability for a monocromatic beam takes, in standard notation, the form

$$P = \sin^2 2\theta \cdot \sin^2 \left(1.27 \frac{\Delta m^2 L}{E}\right) \qquad (1)$$

where $L$ is the neutrino path length (Km), $E$ the beam energy (GeV), and the square mass difference $\Delta m^2$ is measured in eV$^2$.

The sensitivity of a generic experiment to neutrino oscillation is then roughly characterized by the minimum value $\Delta m^2_{min}$ (at $\sin^2 2\theta = 1$) and the minimum value $\sin^2 2\theta_{min}$ (at large $\Delta m^2$) that can be probed within an assigned confidence level (usually the 90% C.L.). This knowledge allows one to draw approximate exclusion plots in the familiar ($\sin^2 2\theta$, $\Delta m^2$) plane, as described, for instance, in Ref. [47].

Following the above reference, one can alternatively use the average value of the baseline/energy ratio $\langle L/E \rangle$ and the minimum detectable oscillation probability $P_{min}$, through the correspondence

$$\sin^2 2\theta_{min} = 2 P_{min} \quad , \quad \Delta m^2_{min} = \frac{\sqrt{P_{min}}}{1.27 \langle L/E \rangle} \quad . \qquad (2)$$

Although this approximate correspondence is not really expected to hold for disappearance searches, it happens to work reasonably well for the cases considered here, when $L$ is taken as the far-detector distance.

In Fig. 1, all the experiments mentioned so far are thus charted in the ($\langle L/E \rangle$, $P_{min}$) plane, as well as in the corresponding ($\Delta m^2_{min}$, $\sin^2 2\theta_{min}$) coordinates, together with some supplementary information: accelerator or reactor source, short or long baseline, oscillation channel(s) probed.

The star labeled ≪ Atmospheric ≫ in Fig. 1 has the coordinates ($\Delta m^2_{min}$, $\sin^2 2\theta_{min}$)= (0.004 eV$^2$, 0.5), which are, approximately, the typical minimum mass/mixing values *allowed* by the atmospheric neutrino data, when the anomaly is interpreted in terms of two-flavor ($\mu \leftrightarrow e$ or $\mu \leftrightarrow \tau$) neutrino oscillations [48]. In this case, however, there is no immediate correspondence with $\langle L/E \rangle$ or $P_{min}$.

Figure 1 contains a great deal of information about the experimental performances, that, however, cannot be fully exploited in a two-generation approach. For instance, at a glance one realizes that future reactor and accelerator experiments will probe values of both $\Delta m^2$ and $\sin^2 2\theta$ lower than those favored by atmospheric neutrino data. Nevertheless, should atmospheric $\nu_\mu$'s oscillate not exclusively into $\nu_e$ or $\nu_\tau$, but in both channels at the same time, one would have no indication how to test this hypothesis completely. In other words, a three-generation approach is needed. An optimal choice should be general enough to cover simultaneously all oscillation channels, and sufficiently simple to get a quick understanding of their interplay. These are the properties of the representations we suggest in the next section.

## 3 Formalism and Graphical Representations

In this Section we recall the basic formalism that will be used throughout the paper, and introduce new suitable representations of the parameter space, together with their application to atmospheric neutrinos.

### 3.1 Basic Notation

The formalism of three-neutrino oscillation in vacuum is very simple. The flavor eigenstates $\nu_e$, $\nu_\mu$ and $\nu_\tau$ can be projected onto the mass eigenstates $\nu_1$, $\nu_2$ and $\nu_3$ (with masses $m_1$, $m_2$ and $m_3$) through a unitary mixing matrix $U$:

$$\nu_\alpha = \sum_{i=1}^{3} U_{\alpha i} \nu_i \quad (\alpha = e, \mu, \tau) \quad , \tag{3}$$

where the $U_{\alpha i}$ can be expressed, as in the quark sector, in terms of three Euler angles $\theta_{ij}$ and a CP-violating phase $\delta$: $U_{\alpha i} = U_{\alpha i}(\theta_{12}, \theta_{23}, \theta_{13}, \delta)$. The standard parametrization of Ref. [47] is adopted.

The indices $i = 1, 2, 3$ can be assigned arbitrarily to the mass eigenstates $\nu_i$, as far as $\theta_{ij} \in [0, \pi/2]$. We overcome here the limiting assumption $\theta_{ij} \in [0, \pi/4]$ of our previous works [13, 14]. All neutrino oscillation effects can then be expressed in terms of six independent variables: $m_2^2 - m_1^2$, $m_3^2 - m_1^2$, $\theta_{12}$, $\theta_{23}$, $\theta_{13}$, and $\delta$.

A less general but more manageable framework is obtained in the limit [6, 7]:

$$|m_2^2 - m_1^2| \ll |m_3^2 - m_{1,2}^2| \equiv m^2 \quad , \tag{4}$$

sometimes called "one mass-scale dominance." In other words, two of the massive states, conventionally identified with $\nu_1$ and $\nu_2$, are assumed to form an almost degenerate doublet $(m_1 \simeq m_2)$ with respect to the mass gap $m$. Accelerator, reactor, and atmospheric neutrinos are meant to probe $m^2$-driven oscillations. Notice that Eq. (4) can hold both for $m_3^2 < m_1^2 \simeq m_2^2$ and for $m_1^2 \simeq m_2^2 < m_3^2$, although only the latter case (corresponding to a natural mass hierarchy) is theoretically appealing, being motivated by the see-saw mechanism [2] and by analogy with the charged fermion masses.

Let us briefly recall the advantages resulting from the approximation in formula (4): *i*) A single (squared) mass scale $m^2$ is involved; *ii*) The angle $\theta_{12}$ can be rotated away by a redefinition of the degenerate states $\nu_1$ and $\nu_2$; and *iii*) CP-violating effects are unobservable [49], so that $\delta$ and the $\nu/\bar{\nu}$ distinction can be discarded, and the elements of $U_{\alpha i}$ taken real. The parameter space is thus reduced to the three variables $(m^2, \theta_{23}, \theta_{13})$. We prefer to adopt the notation introduced in Ref. [50] and that we used in [13, 14]: $\theta_{23} = \psi$, $\theta_{13} = \phi$. In this notation, one has that:

$$U_{e3}^2 = \sin^2\phi \quad , \quad U_{\mu 3}^2 = \cos^2\phi \sin^2\psi \quad , \quad U_{\tau 3}^2 = \cos^2\phi \cos^2\psi \quad . \tag{5}$$

The probability of oscillation $P$, defined as $P = P(\nu_\alpha \to \nu_\beta) \equiv P_{\alpha\beta}$ for appearance experiments, and as $P = 1 - P(\nu_\alpha \to \nu_\alpha) \equiv 1 - P_{\alpha\alpha}$ for disappearance experiments ($\alpha, \beta = e, \mu, \tau$), is then given by:

$$P = 4U_{\alpha 3}^2 U_{\beta 3}^2 S \quad \text{(appearance)} \quad , \tag{6}$$
$$P = 4U_{\alpha 3}^2 (1 - U_{\alpha 3}^2) S \quad \text{(disappearance)} \quad , \tag{7}$$

where $S$ is the oscillation factor $S = \sin^2(1.27 \cdot m^2 L/E)$. Notice that, as far as a single experiment is concerned, the oscillation probability $P$ reduces to the simple two-neutrino form of Eq. (1) through the replacements:

$$\sin^2 2\theta \Leftrightarrow 4U_{\alpha 3}^2 U_{\beta 3}^2 \quad \text{(appearance)} \quad , \tag{8}$$
$$\sin^2 2\theta \Leftrightarrow 4U_{\alpha 3}^2 (1 - U_{\alpha 3}^2) \quad \text{(disappearance)} \quad , \tag{9}$$

and the obvious identification $\Delta m^2 \equiv m^2$. Thus the one mass-scale dominance can be considered as the simplest three-flavor extension of the two-flavor scenario.

In the following, all the experiments will be reanalyzed by using Eq. (6) or (7) instead of Eq. (1), and the combined information will be presented in the $(m^2, \psi, \phi)$ parameter space. We discuss the latter issue in the next subsection, by proposing two new useful graphical representations.

## 3.2 Triangular and Bilogarithmic $(\text{tg}^2\psi, \text{tg}^2\phi)$ Representations

The $(m^2, \psi, \phi)$ parameter space is equivalent to $(m^2, U_{e3}^2, U_{\mu 3}^2, U_{\tau 3}^2)$ with the unitarity constraint $U_{e3}^2 + U_{\mu 3}^2 + U_{\tau 3}^2 = 1$. For any $m^2$, this constraint can be graphically exploited by identifying the $U_{\alpha i}^2$ with the three heights projected from a point inside an equilateral triangle (of unit height) onto the sides, as shown in Fig. 2 [51]. The point in the triangle represents $\nu_3$, and the corners the flavor eigenstates $\nu_{e,\mu,\tau}$, so that when $\nu_3$ coincides with one of the corners, no oscillation occurs. For $\nu_3$ on a side, pure two-flavor oscillations take place. For $\nu_3$ strictly inside the triangle, none of the $U_{\alpha i}^2$ is zero and a true three-flavor oscillation scenario emerges. The position of $\nu_3$ can also be charted by the two independent coordinates $\sin^2\psi$ and $\sin^2\phi$, as shown in the lower part of Fig. 2.

Figure 3 shows curves of iso-probability $P$ in this triangular representation, as derived from Eqs. (6) and (7) for two representative cases ($P = 0.1, 0.3$) in the different oscillation channels experimentally open. The "large $m^2$" limit has been chosen, in order to average $S$ to $1/2$; however, iso-P curves for finite values of $m^2$ would be similar.

The "cross-talk" among the different oscillation channels is neatly and symmetrically displayed in Fig. 3. If all searches find null results, than iso-$P$ lines would represent approximately the boundaries of excluded regions. In this case, a qualitative combination of "$P > 0.1, 0.3$ excluded regions" is shown in the triangle labeled "ALL."

This representation would be very useful to pinpoint the flavor components of $\nu_3$, if they were all large. At present, however, it cannot be excluded that one or both angles $\psi$, $\phi$ are very close to 0 or $\pi/2$. In this case, $\nu_3$ would be very close (graphically indistinguishable) to the triangle boundary. To overcome this potential problem, another representation is introduced.

Actually, a suitable expansion of the parameter region close to the sides of the triangle ($\sin^2\phi$, $\sin^2\psi = 0, 1$) is obtained by adopting the rectangular coordinates $\log \text{tg}^2\psi$ and $\log \text{tg}^2\phi$. This choice "magnifies" the four cases $\psi, \phi = \varepsilon, \pi/2 - \varepsilon$ (for $\varepsilon \to 0$). Notice also that in the small mixing angle regime ($\psi$ and $\phi \to 0$) these coordinates are equivalent, up to terms of $\mathcal{O}(\psi^3, \phi^3)$, to a bilogarithmic ($\sin^2\psi$, $\sin^2\phi$) scale.

In Fig. 4, the same curves of iso-probability of Fig. 3 are shown in this new representation. The price paid to magnify the corners in the $\triangle \to \square$ mapping is the "flavor symmetry breaking" of the plot. However, small oscillation probabilities, like those probed by many of the experiments in Fig. 1, can be easily represented without shrinking the maximal mixing cases (corresponding now to $\log \text{tg}^2\psi$, $\log \text{tg}^2\psi \simeq 0$) too much. The ($\text{tg}^2\psi$, $\text{tg}^2\phi$) plot is thus appropriate to the analysis of accelerator and reactor data of Sec. 4.

### 3.3 Application to Atmospheric Neutrino Oscillations

Here the previous representations are applied to a more complex case, namely, atmospheric neutrino oscillations.

In order to obtain an approximate representation of the parameter space allowed by the atmospheric neutrino anomaly, it is sufficient to consider only the double ratio $R$ of measured (Data) to simulated (MC) $\mu$-like to $e$-like event rates: $R = (\mu/e)_{Data}/(\mu/e)_{MC}$, although we have argued in Ref. [23] that more refined analyses can be done by separating the $\mu$ and $e$ flavor information.

In the presence of oscillations, the expected value of $R$ is approximately given by [9]

$$R \simeq \frac{\langle P_{\mu\mu}\rangle + \langle P_{e\mu}\rangle/r}{\langle P_{ee}\rangle + \langle P_{\mu e}\rangle \cdot r} \quad , \tag{10}$$

where $\langle\rangle$ means the average over neutrino energy, direction, etc., and the numerical value of $r$ is $\sim 2.5$.

For large values of $m^2$, one has roughly $\langle S \rangle \simeq 1/2$, and the corresponding iso-$R$ lines derived through Eq. (10) are represented in Fig. 5 in both the triangular and the $(\text{tg}^2\psi, \text{tg}^2\phi)$ plots. A "typical" anomalous result as $R = 0.5 \pm 0.2$ would favor the region inside the curves at $R = 0.7$. Notice that the usual two-flavor oscillation analyses [48] probe only the lower side ($\mu \leftrightarrow \tau$) or right-hand side ($\mu \leftrightarrow e$). Here, these subcases are smoothly interpolated by a more general $e \leftrightarrow \mu \leftrightarrow \tau$ oscillation solution. The merging of the two-flavor solutions has also been shown, in different forms, in Refs. [10, 12, 15].

Lowering the value of $m^2$ below $10^{-2}$ eV$^2$, the $R \lesssim 0.7$ region gradually shrinks, and disappears for $m^2 \lesssim 4 \times 10^{-3}$ eV$^2$, at essentially the same rate as in the two-flavor case.

Although rather approximate, the above analysis is sufficient for our purposes, as the comparison with long baseline experiments performed in Sec. 6. A more quantitative analysis of all atmospheric neutrino experiments and data will be presented elsewhere [24].

## 4  Analysis of Present Accelerator and Reactor Data

The formalism and the $(\text{tg}^2\psi, \text{tg}^2\phi)$ representation introduced in the previous section are applied here to the main subject of this work: a thorough analysis of the most constraining accelerator and reactor neutrino oscillation (completed) experiments: E776, E531, CDHSW, Krasnoyarsk, Bugey and Gösgen.

Although the combined Bugey and Krasnoyarsk bounds actually improve those from Gösgen alone, we have decided to include this experiment to make the global fit more robust, and get sharper limits in the low $m^2$ range ($m^2 \lesssim 0.1$ eV$^2$) which is (at present) probed only by reactors and is particularly interesting for the comparison with the atmospheric neutrino results.

### 4.1  Data Analysis

In order to draw precise bounds in the new parameter space, and especially to combine all data consistently, we have chosen the hard way of reanalyzing the six experiments E776, E531, CDHSW, Krasnoyarsk, Bugey, and Gösgen, starting from their raw data.

We have processed, for each experiment, as much data as can be recovered from the published papers, including neutrino energy spectra, energy resolution functions, measured and simulated event histograms, efficiencies, backgrounds, statistical and (correlated) systematic errors. In the presence of neutrino oscillations, the smearing effect over the source size has been included. Statistical $\chi^2$ tests have been performed to compare data and predictions for single and combined experiments, in order to assign a well-defined confidence level to any given oscillation scenario. All the ingredients of this analysis are detailed in the Appendix.

An obvious intermediate check is to reproduce the published two-flavor plots. Thus, in Fig. 6 we show the 90% C.L. exclusion curves (solid lines) that we have obtained through a $\chi^2$ analysis of the six experiments above. More precisely, these contours correspond to $\chi^2(\sin^2 2\theta, \Delta m^2) - \chi^2_{min} = 4.61$ (2 d.o.f.). Also shown are the 99% C.L. bounds ($\Delta\chi^2 = 9.21$, dotted lines). The vertical scale $\Delta m^2$ ranges from $2 \times 10$ to $2 \times 10^{-3}$ eV$^2$, that is from the values relevant for the $\nu$ contribution to the dark matter problem, to just below the present sensitivity. The horizontal scale for $\sin^2 2\theta$ is either linear or logarithmic, in order to facilitate the comparison with the original, published 90% C.L. bounds (not shown), a comparison which is very satisfactory in all cases. All these experiments are highly consistent with the no-oscillation scenario: $\chi^2_{no\ osc.} - \chi^2_{min} \lesssim 1$. This will prove useful later.

## 4.2 Separate and Combined Bounds in the $(\text{tg}^2\psi, \text{tg}^2\phi)$ Plane

Having checked the two-flavor limits, let us proceed to the more general three-neutrino analysis in the $(\text{tg}^2\psi, \text{tg}^2\phi)$ plane at fixed $m^2$.

In Fig. 7, the exclusion plots at the 90% and 99% C.L. are shown in the separate oscillation channel, and then combined (ALL), for $m^2$ fixed at 20 eV$^2$. Notice that the final result is dominated by the channels $e \leftrightarrow e$, $e \leftrightarrow \mu$ and $\mu \leftrightarrow \tau$. This is not only related to the fact that different experiments have different sensitivities, but also to the unitarity of the mixing matrix, which implies that the results from different oscillation channels *must* be redundant. By displaying the experimental results as in Fig. 7, the weight of each experimental channel in the fit and the redundance of the information (that would become a consistency check in case of neutrino oscillation evidences) are particularly manifest.

Lowering $m^2$ to 2 eV$^2$ (Fig. 8), the $e \leftrightarrow \tau$ channel does not contribute anymore, as also expected from Fig. 6 (E531). With respect to Fig. 7, the $\mu \leftrightarrow \tau$ ($\mu \leftrightarrow \mu$) excluded region is reduced (enlarged), so that the final results are now dominated by the channels $e \leftrightarrow e$, $e \leftrightarrow \mu$ and $\mu \leftrightarrow \mu$.

For $m^2$ as low as 0.2 eV$^2$ (Fig. 9), only reactor and E776 data are still sensitive to neutrino oscillations (this can also be understood by looking at Fig. 6). Their final combination is dominated by reactor data ($e \leftrightarrow e$ channel) whose sensitivity survives until $m^2 \simeq 7 \times 10^{-3}$ eV$^2$, which is the lowest value probed by the Krasnoyarsk experiment.

The interplay among the different channels in the three-flavor framework being clarified, let us summarize only the final results in Fig. 10, where all data are combined for twelve representative values of $m^2$, ranging from 20 to $5 \times 10^{-3}$ eV$^2$. For the largest values of $m^2$, only the regions corresponding to almost pure flavor eigenstates are allowed. Lowering $m^2$, the limits coming from the $\mu \leftrightarrow \mu$ and $\mu \leftrightarrow \tau$ oscillation channels are weakened, and for $m^2$ between 0.5 and 0.2 eV$^2$ maximal $\mu \leftrightarrow \tau$ mixing (tg$^2\psi = 1$ and tg$^2\phi = 0$) is allowed. At even lower values of $m^2$, the limits imposed by $\mu \leftrightarrow e$ searches become weaker than reactor limits (horizontal stripes), which in any case forbid maximal $e \leftrightarrow \mu$ and $e \leftrightarrow \tau$ mixing. In the last plot ($m^2 = 5 \times 10^{-3}$ eV$^2$) data are insensitive to neutrino oscillations, and any mixing is allowed. We note in passing that, if the two quasi-degenerate states $\nu_1$ and $\nu_2$ were used to solve the solar neutrino problem through matter-enhanced oscillations, than the large-$\phi$ region would be disfavored for any $m^2$ by fits to present solar $\nu$ data [13, 14].

A technical point is in order. As already noticed, all experiments are in good agreement with "no oscillation." Thus, the minima reached in the fits at various fixed values of $m^2$ (Figs. 7–10) are always very close, in a $\chi^2$ sense, to the absolute minimum in the $(m^2, \psi, \phi)$ space, which, by itself, is not of particular interest. Thus, a further minimization with respect to $m^2$ (not shown) would modify the C.L. contours in Figs. 7–10 only slightly. It is intended that this circumstantial shortcut, that avoids minimization over $m^2$, cannot be applied when one or more experiments in the fit do show evidence of neutrino oscillations.

Figure 10 is the main result of our work. It represents a concise and neat summary of the present results from neutrino oscillation searches in the various channels, and shows the usefulness of the proposed framework for phenomenological analyses.

## 4.3 Implications for Atmospheric Neutrinos

Let us discuss qualitatively the implications of the above results on the atmospheric neutrino anomaly. From Figs. 5 and 10, and from the discussion in Sec. 3.3, it is evident that accelerator and reactor bounds completely *exclude* the zone favored by atmospheric data for $m^2 \gtrsim 0.5$ eV$^2$. For a lower $m^2$ ($0.01 \lesssim m^2 \lesssim 0.5$ eV$^2$), the shape of the "atmospheric solution" would still not change very much, while the present bounds from reactors would cut most, or all, of the upper part of this solution. In particular, pure $e \leftrightarrow \mu$ oscillations of atmospheric neutrinos are forbidden for $m^2 \gtrsim 0.02$ eV$^2$. For $m^2$ decreasing from $\sim 0.01$ to $\sim 0.005$ eV$^2$, both the atmospheric favored zone and the reactor bounds would become weaker and eventually vanish.

The possibility of constraining from above the range of $m^2$, by comparing the atmospheric neutrino favored region with the accelerator/reactor data, is a valuable result that deserves a separate quantitative analysis [24]. Earlier discussions can be found in Refs. [13, 14], in the assumption $\psi, \phi \in [0, \pi/4]$, and in Ref. [15], without any limiting assumption on mixing, but also without accelerator data.

# 5 Review of Near and Far Future

In this section we use the well-established accelerator and reactor bounds, the neutrino oscillation solution to the atmospheric neutrino anomaly and the tentative LSND oscillation signal to "review" the possible results that can be expected in the next few years or decade by future experiments. LSND data have interesting implication on short baseline experiments, and atmospheric data mainly on future long baseline experiments.

## 5.1 The LSND Signal and Future Short Baseline Experiments

Recently, the LSND collaboration has reported [3] a possible signal of oscillation in the $\mu \leftrightarrow e$ channel. The corresponding band favored at 95% C.L. in the ($\sin^2 2\theta$, $\Delta m^2$) plane is shown in their Fig. 3. An unofficial reanalysis of the same data [34], which is claimed to be more robust against variations of the selection criteria, is however consistent with no oscillation.

In view of the preliminary status of the LSND data, no attempt to re-fit them is made here, and the above 95% C.L. band is simply mapped in the ($\text{tg}^2\psi$, $\text{tg}^2\phi$) plane with the help of Eq. (8). In Fig. 11, this exercise is repeated for several representative values of $m^2$. Superposed to the LSND *allowed* region (inside solid lines), the region *excluded* at 95% C.L. by all data is also shown (inside dotted lines). There is no (or scarce) compatibility, except for the cases $m^2 = 0.5$, 1 eV$^2$, where two thin (almost horizontal) zones, at small and large values of $\phi$, survive the comparison. A more dense scan of $m^2$ values shows that a reasonable compatibility interval is $0.3 \lesssim m^2 \lesssim 2$ eV$^2$.

A similar conclusion on the $m^2$ range has been reached in Ref. [3], although with a weaker argument (the $\tau$ flavor was discarded and the $e \leftrightarrow e$, $\mu \leftrightarrow e$ bounds were superposed in the same ($\sin^2 2\theta$, $\Delta m^2$) two-flavor plane). Here the combined information on $\psi$ and $\phi$ shows that a substantial mixing with the $\tau$ flavor is also allowed.

It is interesting to check if the large $\tau$-mixing part of the above compatibilty region can be probed, in the $\mu \leftrightarrow \tau$ channel, by the CERN experiments CHORUS and NOMAD. In Fig. 12 we thus show the sensitivity region of CHORUS+NOMAD (below dotted lines), as derived by the corresponding ($\langle L/E \rangle$, $P_{min}$) values in Fig. 1. It is evident that the two CERN experiments can probe a non-negligible fraction of the LSND signal for $m^2 \gtrsim 0.5$ eV$^2$. Unfortunately, a comparison with Fig. 11 shows that the testable zone is already excluded by all present data at any $m^2$. No "LSND-induced" signal is thus expected in the $\mu \leftrightarrow \tau$ channel probed by CHORUS or NOMAD.

Should a signal be found at CERN and the LSND evidence also confirmed, the one mass-scale approximation would be in question, and more complicated oscillation frameworks, perhaps with two comparable mass differences, would be needed to accomodate both negative and positive searches.

Concerning the planned COSMOS experiments in the $\mu \leftrightarrow \tau$ channel, the expected sensitivity regions would be similar to those of CHORUS+NOMAD in Fig. 12, but en-larged by one decade, and thus potentially able to probe a small part of the LSND solution not yet excluded by present data. However, well before COSMOS operation, the running KARMEN experiment ($\mu \leftrightarrow e$) is expected to check completely, and independently, the same parameter region now explored by LSND. Future short baseline (reactor, $e \leftrightarrow e$) experiments as Chooz and San Onofre can also completely (dis)prove the LSND claim, by placing bounds in large horizontal stripes of the ($tg^2\psi$, $tg^2\phi$) plane.

## 5.2 Atmospheric Neutrino Anomaly and Future Long Baseline Experiments

One of the most important issues that long baseline neutrino experiments are meant to clarify is the atmospheric neutrino anomaly. The information contained in Fig. 1 about the experiments, and the oscillation solution to the anomaly discussed in Sec. 3 are sufficient to understand the basic aspects of their interplay.

In Fig. 13, the sensitivity regions of some short/long baseline experiments are shown (solid lines), together with the approximate region favored by the anomalous atmospheric neutrino data (dotted lines). A reference value $m^2 = 0.01$ eV$^2$ has been chosen, which is close to typical, atmospheric best-fit values of $m^2$ [48], and for which the present reactor data still leave a large part of the ($tg^2\psi$, $tg^2\phi$) plane to be explored (see Fig. 10); however, many of our conclusions are independent of specific values of $m^2$.

In Fig. 13 it is shown that a single $\mu \leftrightarrow \mu$ disappearance experiment, as BNL E889, is able to probe completely the region favored by the anomaly. Another exhaustive test is the combination of the appearance channels $\mu \leftrightarrow e$ and $\mu \leftrightarrow \tau$, as in the CERN to ICARUS proposal. In this case, the region where $\nu_3$ has large components on all the three flavor eigenstates would be tested twice. Redundant covering of the atmospheric neutrino region is also obtained by joining $\mu \leftrightarrow e$ to $\mu \leftrightarrow \mu$ searches, as in KEK to SuperKamiokande, and even more by testing all three channels involving $\mu$-flavor, as in the MINOS (and/or P-822) proposals. The redundancy could be used for useful cross-checks of possible oscillation signals. Insufficient covering is obtained in $e \leftrightarrow e$ disappearance searches (Chooz, San Onofre), or in appearance searches involving only one channel, as for CERN to NESTOR ($\mu \leftrightarrow e$).

Fig. 13 is a good basis to understand what happens if a signal is found. For instance, a signal in the $\mu \leftrightarrow \tau$ channel in ICARUS would isolate a ∩-shaped band. A signal in Chooz or San Onofre would instead isolate a horizontal stripe. The band-stripe intersection would determine two spots in the ($tg^2\psi$, $tg^2\phi$) plane, and the ambiguity should be solved by a third signal in a different channel. A similar situation would show up if signals were found in both $\mu \leftrightarrow e$ and $\mu \leftrightarrow \tau$ channels: a third experiment would be required to choose between the two solutions (a requirement overlooked in Ref. [20]). Experiments performing *three* independent searches, like MINOS, can thus potentially isolate a single solution in the ($tg^2\psi$, $tg^2\phi$) plane. Of course, determining the allowed range of $m^2$ would require additional spectral analyses of the event samples, that seem to be experimentally feasible only for relatively large mixing cases (see the notable discussion in the MINOS proposal [44]).

In conclusion, as far as the atmospheric neutrino anomaly is taken as an indicator for the discovery potential of long baseline experiments, an analysis like the one presented in Fig. 13 allows one to understand clearly the interplay between the various channels explored, and to judge the ability of single and combined experiments to solve this issue.

# 6 Summary and Conclusions

We have presented in this paper a realization of the one mass-scale limit of the three-flavor mixing scheme. In this framework, an exhaustive analysis of the most constraining accelerator and reactor data has been performed. The results have been represented in a new and useful graphical form, that shows at a glance the interplay between the different oscillation channels. Implications on the oscillation solution to the atmospheric neutrino anomaly have been briefly discussed, and are worth further study.

A compatibility check of the tentative LSND evidence for oscillations with all data has been made, which shows that only a very limited part of the LSND signal is allowed in the parameter space. This region will be explored in different ways by future short baseline searches, but is beyond the discovery potential of the running CHORUS and NOMAD experiments.

The ability of future short and especially long baseline experiments to cover the mass/mixings space favored by atmospheric neutrinos has also been discussed. Pros and cons of searches in the different channels, as well as consistency checks of hypothetical future signals, have been pointed out, and search strategies sketched.

A great amount of information has been processed in this work, and the analysis of present accelerator and reactor bounds has been particularly painful. Nevertheless, we think that the results have been put in a form that is easy to understand. Our best hope is that, in the analysis of future experiments, thinking in this three-generation framework will be as easy as understanding a familiar two-generation plot.

# Acknowledgements

The authors thank Dr. G. Zacek for useful information about the Gösgen experiment. The work of E.L. is supported in part by a post-doctoral INFN fellowship and by funds of the Institute for Advanced Study.

This research was in part performed under the auspices of the Theoretical Astroparticle Network, under contract N. CHRX-CT93-0120 of the Direction General 12 of the E.E.C.

# Appendix

In this Appendix, our reanalysis of the raw accelerator and reactor data—which is the basis of the results in Sec. 4—is documented for each experiment separately.

**BNL E776** ($\mu \leftrightarrow e$). The E776 experiment [26] searched for $\nu_e$ ($\bar{\nu}_e$) appearance in a wide-band $\nu_\mu$ ($\bar{\nu}_\mu$) beam. The energy distributions of (anti)neutrino-induced events were found consistent with the estimated background.

In our analysis, the published (unoscillated) Monte Carlo histograms have been convoluted with the oscillation probability, taking into account the (anti)neutrino energy spectra, the detector energy resolution function [52] and a smearing over the production region. The results have been compared with the data histograms through a $\chi^2$ statistic appropriate to low-rate bins [53]. As in Ref. [26], $\chi^2$ has been minimized within the background uncertainties.

**FNAL E531** ($\mu, e \leftrightarrow \tau$). The E531 experiment at Fermilab [27] searched for $\tau$ appearance in a (mainly) $\mu$-flavored wide-band beam. The $e$-flavor contamination was considered sufficiently well-understood to put limits in the $e \to \tau$ channel as well. No $\tau$-events were found in either case.

The expected number of events $n_\tau$ is obtained by folding the published neutrino $L/E$ spectra with the overall efficiencies [27] and the oscillation probability. The smearing effect due to the size of the neutrino production region is included. The C.L. contours in Fig. 6 are obtained by using for $n_\tau$ a Poisson statistic with zero mean.

In order to combine the E531 results with those from the other experiments in a global $\chi^2$-analysis, a "fake" $\chi^2$ statistic has been finally determined by probability inversion $[\chi^2 = \chi^2(P)]$, yielding for any given C.L. the same curves as the original Poisson statistic.

**CERN CDHSW** ($\mu \leftrightarrow \mu$). The CDHSW experiment at CERN [28] searched for $\mu \to \mu$ disappearance by means of two (so-called front and back) detectors downstream an unfocused beam. The back/front ratio of muon events, normalized to the squared distance and the target mass ratio, was reported in several bins of "longitudinally projected muon range in iron," which is a measure of the parent neutrino energy. The expected back/front ratio, equal to 1 in the absence of oscillations, was corrected to take into account secondary hadron production and decay, as well as geometrical acceptance effects [28].

We have recalculated the expected back and front muon event rate, by folding the oscillation probabilities at the two sites with the neutrino fluxes, taking into account the peculiar (non-uniform) path-length smearing effect due to the sizable length of the decay tunnel. The correction factors mentioned above are then included in each back/front-ratio bin. The comparison with data is made through a $\chi^2$ statistic, including the correlation of common systematics (dead-time, efficiency uncertainties) in the covariance matrix.

There is a slight disagreement between our 90% C.L. contour in Fig. 6 and the published one [28] around $\Delta m^2 \simeq 10$ eV$^2$. It can be traced back to the fact that in Ref. [28] the $\chi^2$ is minimized with respect to $\sin^2 2\theta$ at any *fixed* $\Delta m^2$, and not with respect to both variables jointly, as has been done in our Fig. 6.

Finally, we note that the CDHSW bounds become rapidly weak for $\Delta m^2 \gtrsim 10$ eV$^2$. Above $\sim 20$ eV$^2$, they would be superseded by the limits obtained in the CCFR experiment at FNAL [54] (not shown in Fig. 6).

**Krasnoyarsk reactors** ($e \leftrightarrow e$). The Russian experiment at Krasnoyarsk [29] searched for $\bar{\nu}_e$ disappearance by means of a detector placed at a distance of 57, 57.6 and 231.4 m from three nuclear reactors. Detection was based on the inverse-$\beta$ process. The reactors were switched on and off in various combinations, allowing an overconstrained determination of both the signal and the background. The data from the two nearest reactors were reduced to one "effective reactor" after a correction ($\sim 1\%$) for the small difference in the distance.

In order to get the positron spectra in the presence of oscillations, we have convoluted the disappearance probability with the $^{235}$U antineutrino spectrum [55], the radiatively-corrected cross section [56] and a smearing function over the reactor (presumable) core size. In the calculation of the expected positron yield, we have also used additional information on some detector parameters (target mass, efficiencies) as reported in [57]. The $\chi^2$ analysis includes the correlation of common systematics: efficiency, reactor power, cross-section, fuel composition uncertainties.

**Bugey reactor** ($e \leftrightarrow e$). In the recent experiment at Bugey [30], a search has been made for $\bar{\nu}_e$ disappearance at three distances (15, 40 and 95 m) from the reactor. Both experimental and simulated (unoscillated) positron spectra at the three distances can be recovered from Ref. [30], together with various estimates of the background (subtracted) and of systematic effects (neutrino flux, cross-section, efficiency uncertainties). Two neutrino oscillation analyses are performed in [30]. In the first (second), the data at the nearest detector is (not) normalized out. We have reproduced the—more constraining—second analysis.

In the presence of neutrino oscillations (smeared over the presumable reactor core size), the positron spectra are recalculated and compared with the data histograms through a $\chi^2$ statistic, including the correlation of common systematic uncertainties. A minimization is performed on one variable, which parametrizes a possible energy bias in the positron spectra (variable "b" in Ref. [30]). We have checked that our analysis is not significantly improved by minimizing with respect to additional variables (five are used in the rather intricated $\chi^2$ function of Ref. [30]). The rich structure of the 90% C.L. contour in Fig. 6 compares well with the published bounds.

**Gösgen reactor** ($e \leftrightarrow e$). The experiment at the Gösgen reactor [31] searched for $e \to e$ disappearance by measuring the positron energy spectra (from the inverse-$\beta$ process) at three distances (37.9, 45.9 and 64.7 m) from the source.

The very detailed Refs. [31, 58] make it possible to recover not only the measured and simulated spectra, but also the (energy-dependent) resolution and efficiency functions, the size of the reactor core and a careful estimate of the errors from various sources. All these ingredients have been used to recalculate the expected positron spectra in the presence of oscillations, and to compare them with the data histograms through a $\chi^2$ statistic. The correlations of common systematics (neutrino spectrum normalization, cross-section, efficiency and reactor power uncertainties) are included. The results shown in Fig. 6 should be compared with the corresponding "Analysis B" plot in Ref. [31].

# Figure Captions

**Fig. 1:** Sensitivity of present and future accelerator and reactor experiments to neutrino oscillations, charted in the ($\langle L/E \rangle$, $P_{min}$) and ($\Delta m^2_{min}$, $\sin^2 2\theta_{min}$) coordinates. The initial ($\alpha$, prepared) and final ($\beta$, detected) neutrino flavors are also indicated ($\alpha \to \beta$). See the text for details.

**Fig. 2:** Triangular plot representation of the neutrino state $\nu_3$, in terms of its flavor components $\nu_e$, $\nu_\mu$, $\nu_\tau$. For a triangle of unit height, the unitarity relation $U^2_{e3} + U^2_{\mu 3} + U^2_{\tau 3} = 1$ is enforced by identifying the $U^2_{\alpha 3}$ with the three distances from $\nu_3$ to the sides (dashed lines). The "general $3\nu$ oscillation" scenario and its "$2\nu$ oscillation" and "no oscillation" limits are represented respectively by the *region inside*, by the *sides*, and by the *corners* of the triangle. Iso-lines of $\sin^2 \psi$ and $\sin^2 \phi$ (parametrizing the elements $U_{\alpha 3}$) are charted in the lower plot.

**Fig. 3:** Iso-lines of oscillation probability $P$ ($P = P_{\alpha\beta}$ or $P = 1 - P_{\alpha\alpha}$) in the triangular plot defined in Fig. 2, for various channels ($\alpha \leftrightarrow \beta$) and large $m^2$, such that $\sin^2(1.27 \cdot m^2 L/E) \simeq 1/2$. In the case of no oscillation, iso-$P$ lines delimit excluded regions. A qualitative combination of null results in all channels is shown in the plot labeled "ALL," where the corners belong to the allowed zone.

**Fig. 4:** As in Fig. 3, but in the bilogarithmic ($\text{tg}^2 \psi$, $\text{tg}^2 \phi$) representation. Notice the deformation of the iso-$P$ contours due to the $\triangle \to \square$ mapping. Here the $\nu_e$, $\nu_\mu$ and $\nu_\tau$ flavor eigenstates are reached respectively in the limits: $\text{tg}^2 \phi \to \infty$ (any $\text{tg}^2 \psi$); $\text{tg}^2 \phi \to 0$ and $\text{tg}^2 \psi \to \infty$; $\text{tg}^2 \phi \to 0$ and $\text{tg}^2 \psi \to 0$.

**Fig. 5:** Curves of iso-$R$ (double flavor ratio) for atmospheric neutrinos, at large $m^2$ (i.e., $\sin^2(1.27 \cdot m^2 L/E) \simeq 1/2$), in both the triangular and the ($\text{tg}^2 \psi$, $\text{tg}^2 \phi$) representation. A "typical" experimental result as $R = 0.5 \pm 0.2$ would correspond to the allowed region inside the two $R = 0.7$ lines. Notice that the pure two-oscillation solutions $e \leftrightarrow \mu$ (right-hand side) and $\mu \leftrightarrow \tau$ (bottom side) are smoothly connected here. See the text for the case $m^2 \lesssim 0.01$ eV$^2$.

**Fig. 6:** Two-flavor exclusion plots at the 90% and 99% C.L. for the Gösgen, Bugey, Krasnoyarsk, E531, CDHSW and E776 experiments, as derived by our reanalysis of the raw data. The solid contours agree well with the original 90% C.L. bounds published in the quoted references of the different experiments.

**Fig. 7:** Three-flavor exclusion plots at the 90% and 99% C.L. in the ($\text{tg}^2 \psi$, $\text{tg}^2 \phi$) representation, as derived by the experimental data analyzed in the various oscillation channels ($\alpha \leftrightarrow \beta$), and in all channels combined (ALL) for $m^2 = 20$ eV$^2$.

**Fig. 8:** As in Fig. 7, but for $m^2 = 2$ eV$^2$. Notice that no bound is placed by the E531 experiment ($e \leftrightarrow \tau$ channel) in this case.

**Fig. 9:** As in Fig. 7, but for $m^2 = 0.2$ eV$^2$. Notice that only reactor and E776 data give constraints in this case.

**Fig. 10:** Bounds in the $(\text{tg}^2\psi, \text{tg}^2\phi, m^2)$ parameter space, shown as $(\text{tg}^2\psi, \text{tg}^2\phi)$ sections at different values of $m^2$, as derived by a reanalysis of all accelerator and reactor data. Only reactor bounds (horizontal stripes) survive at low $m^2$ (they vanish at $m^2 \lesssim 0.007$ eV$^2$).

**Fig. 11:** LSND against Rest of the World. Solid lines delimit the region *allowed* at the 95% C.L. by the LSND signal, as read from Fig. 3 in Ref. [3], and then mapped onto the $(\text{tg}^2\psi, \text{tg}^2\phi)$ plane at different values of $m^2$. Dotted lines delimit the region *excluded* at the same C.L. by all the other experiments, as derived by our analysis. Notice that there is very poor (or no) compatibility, except for the cases $m^2 = 0.5$ and 1 eV$^2$ (more generally, for $0.3 \lesssim m^2 \lesssim 2$ eV$^2$).

**Fig. 12:** LSND vs. CHORUS+NOMAD, for different values of $m^2$. The CHORUS+NOMAD sensitivity region (below the dotted curve) and the LSND signal (inside the solid lines) overlap partially. However, comparison with Fig. 11 excludes the common region in any case: no "LSND-induced" signal is thus expected in CHORUS or NOMAD.

**Fig. 13:** Comparison, at $m^2 = 0.01$ eV$^2$, between the region qualitatively favored by the atmospheric neutrino data (dotted lines), and the sensitivity region of future short and long baseline experiments (solid lines). Notice that the neutrino oscillation solution to the atmospheric anomaly can be completely probed by future $\mu \to \mu$ disappearance experiments (e.g., BNL E889), as well as by the combined $\mu \to e$ and $\mu \to \tau$ appearance searches (as in the "CERN to ICARUS at G.S." proposal).